\documentclass[twocolumn,epjc3]{svjour3}

\usepackage{graphicx}
\usepackage{color}
\usepackage{amssymb}
\usepackage{amsmath}
\usepackage{braket}
\usepackage{bm}
\usepackage{multirow}

\newcommand{\nc}{\newcommand}           
\nc{\vc}[1]     {\mbox{\boldmath $#1$}} 
\nc{\mapleft}[1]{                       
 \smash{\mathop{                        %
  \hbox to 0.90cm{\rightarrowfill} }\limits_{#1}}}
\nc{\figwidth}{0.8}                    

\nc{\mydraft}	{\setlength{\topmargin}{-1.5cm}}
\mydraft   

\begin{document}
\title{Microscopic calculations of $^6$He and $^6$Li with real-time evolution method}

\author{Q. Zhao\thanksref{e1,addr1} 
        \and
        B. Zhou\thanksref{addr2,addr3}
        \and
        M. Kimura\thanksref{addr2,addr1,addr4}
	\and
	H. Motoki\thanksref{addr2}
	\and 
	Seung-heon Shin\thanksref{addr2}
	}

\thankstext{e1}{e-mail: zhao@nucl.sci.hokdai.ac.jp}
	
\institute{Nuclear Reaction Data Centre (JCPRG), Hokkaido University, Sapporo 060-0810, Japan\label{addr1}
           \and
           Department of Physics, Hokkaido University, Sapporo 060-0810, Japan\label{addr2}
           \and
           Institute of Modern Physics, Fudan University, Shanghai 200433, China\label{addr3}
           \and
	   Research Center for Nuclear Physics (RCNP), Osaka University, Ibaraki 567-0047, Japan\label{addr4}
}

\titlerunning{Microscopic calculations of $^6$He and $^6$Li with real-time evolution
method}\authorrunning{Qing Zhao et al.} 

\maketitle
\abstract {The low-lying cluster states of $^6$He ($\alpha$+n+n) and $^6$Li ($\alpha$+n+p) are
calculated by the real-time evolution method (REM) which generates basis wave functions
for the generator coordinate method (GCM) from the equation of motion of Gaussian wave packets. The
$0^+$ state of $^6$He as well as the $1^+$, $0^+$ and $3^+$ states of $^6$Li are calculated as a
benchmark. We also calculate the root-mean-square (r.m.s.) radii of the point matter, the point
proton, and the point neutron of these states, particularly for the study of the halo characters of
these two nuclei. It is shown that REM can be one constructive way for generating effective basis
wave functions in GCM calculations.  
}

\section{Introduction}
The light nuclei have been studied within the view of the cluster feature for more than five
decades~\cite{Brink1966,Fujiwara1980,Tohsaki2001}, and various nuclear theories have been developed
for the study of nuclear clustering~\cite{Oertzen2006,Horiuchi1991,Ito2012}. 
By assuming the cluster structure, various cluster states of light nuclei have been investigated
explicitly~\cite{Lyu2016,Suhara2010,Zhou2012}. However, as the number of the constituent clusters
and nucleons increases or nuclear system becomes dilute, the number of required basis wave functions
increases very quickly. Therefore, a method which can efficiently sift out the basis is highly
desired. For this purpose, many efforts have been made, such as the stochastic
sampling~\cite{Suzuki1998,Itagaki2003,Mitroy2013} and the imaginary-time development
method~\cite{Fukuoka2013}.   

Recently, a newly time-dependent many-body theory has been developed in
Refs.~\cite{Imai2019,Zhou2020,Shin2021} for the calculations of Be and C isotopes. This real-time
evolution method (REM) generates the basis wave function using the equation of motion (EOM)
which has been applied in the study of heavy-ion collisions~\cite{Zhang2011,Ono1996,Ono2019}, but
now is found to be very  effective in searching the basis wave functions for the microscopic
calculations because of its ergodic nature.

We intend to apply the REM on the $0^+$ ground state of $^6$He nucleus ($\alpha$+n+n). It is
a Borromean nucleus consisting of loosely bound and spatially extended three-body systems, typically
composed of a compact core plus two weakly bound neutrons
(n+n+core)~\cite{Ogan1999,Danilin1998,Zhukov1993}. These properties 
can lead to the huge computational difficulties despite of its simple physical structure. Meanwhile,
the low lying states of $^6$Li ($\alpha$+n+p) also can be a good comparison, where the more compact
states ($1^+$ and $3^+$ states with $T=0$) and the dilute state ($0^+$ state with $T=1$) present
simultaneously. In this study, aiming to explore the applicability of REM, we will calculate the
$0^+$ ground state of $^6$He as well as the low lying states of $^6$Li to reproduce the halo and
un-halo properties of these states. 

This paper is organized as follows: Section~\ref{sec:wavefunction} explains the framework of the
wave function and the real-time evolution method (REM). The numerical results including the energy
and the root-mean-square (r.m.s.) radius are presented and discussed in Sec.~\ref{sec:results}. The
conclusion is summarized in Sec.~\ref{sec:conclusion}. 

\section{Framework}
\label{sec:wavefunction}

\subsection{Hamiltonian}

We begin with the Hamiltonian given below
\begin{equation}
\hat{H}=\sum_{i=1}^A \hat{t}_i - \hat{T}_{c.m.} + \sum_{i<j}^A \hat{v}_N + \sum_{i<j}^A \hat{v}_{C} + \sum_{i<j}^A \hat{v}_{LS}
\end{equation}
where $\hat{t}_i$ and $\hat{T}_{c.m.}$ denote the kinetic energies of each nucleon and the center of mass, respectively. $\hat{v}_N$ denotes the effective nucleon-nucleon interaction and $\hat{v}_C$ denotes the Coulomb interaction. $\hat{v}_{LS}$ denotes the spin-orbit interaction.

For the nucleon-nucleon interaction, we take the Volkov No.2 interaction~\cite{volkov1965} as
\begin{equation}
\begin{split}
V(\mathbf{r})=&(W - M\hat{P}^\sigma \hat{P}^\tau + B\hat{P}^\sigma - H\hat{P}^\tau)\\
&\times [V_1\text{exp}(-r^2/c_1^2)+V_2\text{exp}(-r^2/c_2^2)] ~.
\end{split}
\end{equation}
The corresponding exchange parameters are, $W=0.4$, $M=0.6$ and $B=H=0.125$. The parameters in the
Gaussian terms are, $V_1 = -60.65$ MeV, $V_2 = 61.14$ MeV, $c_1 = 1.80$ fm and $c_2 = 1.01$ fm.   

We take the G3RS potential~\cite{Tamagaki1968,Yamaguchi1979} as the spin-orbit interaction, 
\begin{equation}
V_{ls} = V_0(e^{d_1 r^2} - e^{d_2 r^2})\hat{P}_{31}\hat{L}\cdot \hat{S}~.
\end{equation}
The strength parameter $V_0$ is set to be 2000 MeV. The Gaussian parameters $d_1$ and $d_2$ are set
to be $5.0$ fm$^{-2}$ and $2.778$ fm$^{-2}$, respectively. 

\subsection{Generator coordinate method}

In the current work, the single-particle wave function $\phi(\mathbf{r},Z)$ are expressed
in a Gaussian form multiplied by the spin-isospin part $\chi_{\tau,\sigma}$ as 
\begin{equation}
\phi(\mathbf{r},Z) =
 (\frac{2\nu}{\pi})^{3/4}\text{exp}[-\nu(\mathbf{r}-\frac{\bm{z}}{\sqrt{\nu}})^2+\frac{1}{2}z^2]\chi_{\tau,\sigma}. 
\end{equation}
Here the coordinate $Z$ represents the generator coordinates, which includes the three-dimensional
coordinate $\bm{z}$ for the spatial part of the wave function as well as the spinor $a$ and $b$ for
the spin part $\chi_\sigma = a\ket{\uparrow}+b\ket{\downarrow}$. In this work, the spinor $a$ and
$b$ are also regarded as time-dependent variables which will be generated similarly to the spatial
coordinates as introduced later. The harmonic oscillator parameter $b=\sqrt{1/(2\nu)}=1.46$ fm,
which is same with that used in Refs.~\cite{Itagaki2003,Furumoto2018}. 

We describe the $^6$He and $^6$Li as the $\alpha$-cluster plus two valence nucleon systems in 
the wave function. Thus the corresponding wave function can be written as 
\begin{equation}
\Phi(Z_1,Z_2,\bm{z}_\alpha)=\mathcal{A}\{\phi(\mathbf{r}_1,Z_1)\phi(\mathbf{r}_2,Z_2)\Phi_\alpha(\mathbf{r}_{3-6},\bm{z}_\alpha)\}~. 
\end{equation}
Here $\Phi_\alpha$ is the wave function of the $\alpha$-cluster with $(0s)^4$ configuration. $\phi$
are the single-particle wave functions as introduced above, which are used to describe the valence
nucleons in $^6$He and $^6$Li. Thus, the coordinates $\mathbf{r}_{1}$ and $\mathbf{r}_{2}$ represent
the real spatial position of valence nucleons while $\mathbf{r}_{3-6}$ are for the nucleons in the
$\alpha$-cluster.  

Within the framework of generator coordinate method (GCM), the final wave function is the
superposition of the basis wave functions with different sets of generator coordinates
($Z_1,Z_2,\bm{z}_\alpha$): 
\begin{equation}
\Psi = \sum_i f_i \hat{P}^{J^\pi}_{MK}\Phi_i(Z_{1,i},Z_{2,i},\bm{z}_{\alpha,i})
\end{equation}
where $\hat{P}^{J^\pi}_{MK}$ is the parity and the angular momentum projector. The generator
coordinates $\mathbf{Z}$ can be obtained by solving the equation of motion in REM as explained in
the next subsection. The corresponding coefficients $f_i$ will be determined by the diagonalization
of the Hamiltonian.  

\subsection{Real-time evolution method}

In the quantum system, the wave function should satisfy the Schrodinger equation at all times. Thus, the time-dependent variational principle holds for the intrinsic wave function mathematically:
\begin{equation}
\label{eq:timVarPrin}
\delta\int dt \frac{\bra{\Phi(\mathbf{Z}_1,\mathbf{Z}_2,\mathbf{Z}_\alpha)}i\hbar~d/dt - \hat{H} \ket{\Phi(\mathbf{Z}_1,\mathbf{Z}_2,\mathbf{Z}_\alpha)}}{\langle\Phi(\mathbf{Z}_1,\mathbf{Z}_2,\mathbf{Z}_\alpha)|\Phi(\mathbf{Z}_1,\mathbf{Z}_2,\mathbf{Z}_\alpha)\rangle} = 0
\end{equation}
Regarding the coordinate $Z$ as the function of the time  $t$, we obtain the
equation of the motion (EOM) as
\begin{equation}
\label{eq:EOM1}
i\hbar\sum_{j=1,2,\alpha}\sum_{\sigma=x,y,z,a} C_{i\rho
j\sigma}\frac{dZ_{j\sigma}}{dt}=\frac{\partial \mathcal{H}_\text{int}}{\partial Z^*_{i\rho}} 
\end{equation}

\begin{equation}
\label{eq:EOM2}
\mathcal{H}_\text{int}\equiv\frac{\bra{\Phi(\mathbf{Z}_1,\mathbf{Z}_2,\mathbf{Z}_\alpha)}\hat{H}\ket{\Phi(\mathbf{Z}_1,\mathbf{Z}_2,\mathbf{Z}_\alpha)}}{\langle\Phi(\mathbf{Z}_1,\mathbf{Z}_2,\mathbf{Z}_\alpha)|\Phi(\mathbf{Z}_1,\mathbf{Z}_2,\mathbf{Z}_\alpha)\rangle}
\end{equation}

\begin{equation}
\label{eq:EOM3}
C_{i\rho j\sigma}\equiv\frac{\partial^2 \text{ln}\langle\Phi(\mathbf{Z}_1,\mathbf{Z}_2,\mathbf{Z}_\alpha)|\Phi(\mathbf{Z}_1,\mathbf{Z}_2,\mathbf{Z}_\alpha)\rangle}{\partial Z^*_{i\rho}\partial Z_{j\sigma}}
\end{equation}
By following the EOM, from an initial wave function at $t=0$, the sets of the generator coordinates
($Z_1,Z_2,\bm{z}_\alpha$) for GCM can be yielded as a function of time $t$. The ensemble of the
basis wave functions $\Phi_i(Z_{1,i},Z_{2,i},\bm{z}_{\alpha,i})$ denoted by these sets of
the generated coordinates will hold the information of the quantum system. Thus, effective basis can
be generated. 

In practical calculations, we choose the proper initial excitation energy (the definition can be
found in Ref.~\cite{Imai2019,Zhou2020}) for obtaining various cluster configurations in the
evolution. To avoid the clusters or valence nucleons move to unphysical regions, the rebound
condition is imposed in our REM calculations. By following the work in Ref.~\cite{Ono1996}, we add a 
potential barrier to the Hamiltonian during the REM procedure with the form: 
\begin{equation}
\begin{split}
V_\text{reb}=\frac{k}{2}\sum_i &f(|\mathbf{R}_i-\mathbf{R}_\text{c.m.}|)\\
                         f(x) = (x&-d)^2\theta(x-d)                    \\
\mathbf{R}_i=\frac{\text{Re}(\bm{z}_i)}{\sqrt{\nu}},~&\mathbf{R}_\text{c.m.}=\frac{4}{6}~\mathbf{R}_\alpha+\frac{1}{6}\sum_{j=1}^2  
 \mathbf{R}_j~. 
\end{split}
\end{equation}
Here $\mathbf{R}_i$ and $\mathbf{R}_\text{c.m.}$ represent the spatial position of the $i$th valence
nucleon and the center of mass, respectively, so that $|\mathbf{R}_i-\mathbf{R}_{c.m.}|$ is the
distance between them. Because of the step function $\theta(x-d)$, the evolving valence nucleon will
face potential barrier when it is $d$ fm far from the center of mass, and be
smoothly pushed back in later evolution. We set the strength of the potential barrier $k=6$
MeV/fm$^2$, which determines how rapidly the height of the barrier increases. This value is not
physically important as long as it is not too large or too small. The rebound radius parameter $d$
is set to be $8$ fm in our calculations, which is large enough for the current work. 

We perform the above REM process for the intrinsic wave function of $^6$He and obtain an ensemble of
basis. This ensemble of basis are used for both the calculations of $^6$He and $^6$Li. 

\section{Results}\label{sec:results}
\begin{figure}[htbp]
  \begin{center}
   \includegraphics[width=0.9\hsize]{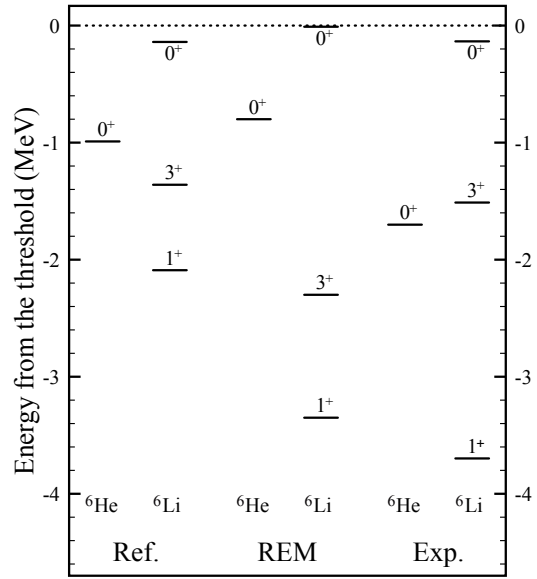}
  \caption{\label{fig:enespec} The energy spectra of $^6$He and $^6$Li. Ref. denotes the results
   from the reference works~\cite{Itagaki2003,Furumoto2018}. REM denotes the results from the
   current work. Exp. denotes the corresponding experimental data~\cite{Tilley2004}. The energy
   is measured relative to the $\alpha$ + n + n threshold, which is set as 0 level with the dotted
   line. For the calculated results of both in this work and the reference works, the energy of
   $^4$He is $-27.57$ MeV, while it is $-28.30$ MeV in the experimental data.} 
  \end{center}
\end{figure}

We firstly show the energy spectra for the low-lying states of $^6$He and $^6$Li nuclei in
Fig.~\ref{fig:enespec}. 
\begin{figure*}[htbp]
\begin{center}
   \includegraphics[width=0.6\hsize]{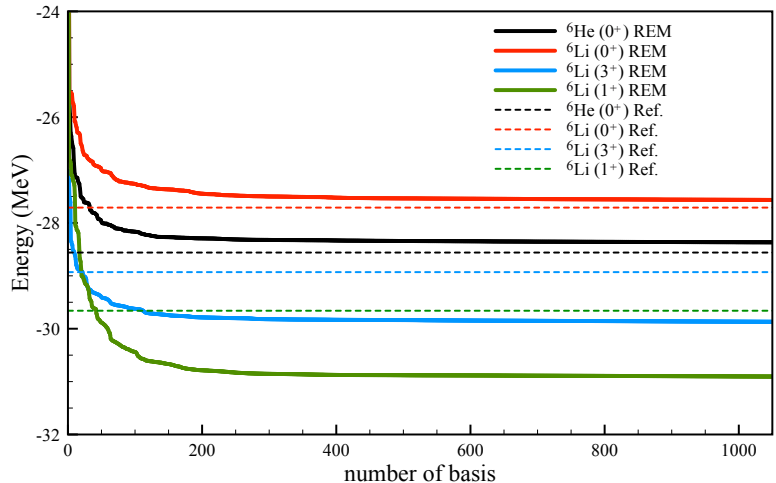}
   \caption{\label{fig:enecon} The energy convergence of $^6$He and $^6$Li from the REM calculations
 concerning the successive addition of bases. The dash lines are the corresponding results from the
 reference works~\cite{Itagaki2003,Furumoto2018}.} 
  \end{center}
\end{figure*}
\begin{figure*}[htbp]
  \begin{center}
   \includegraphics[width=0.7\hsize]{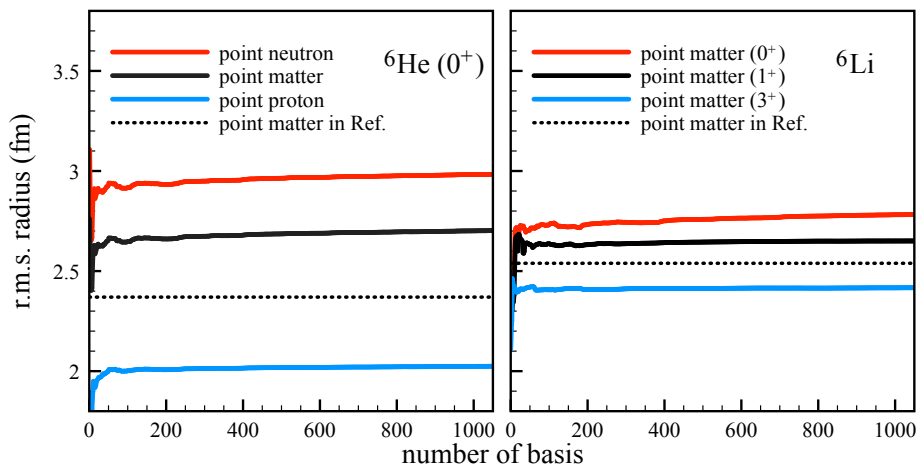}
  \caption{\label{fig:rmscon}The r.m.s. radii of $^6$He and $^6$Li from the REM calculations
   concerning the successive addition of bases. The dotted lines denote the results in the reference
   works~\cite{Itagaki2003,Furumoto2018}.} 
  \end{center}
\end{figure*}
\begin{table*}[htbp]
  \begin{center}
    \caption{The numerical results of the $0^+$ ground state of $^6$He, as well as the $1^+$, $3^+$
   and $0^+$ states of $^6$Li from the calculations of REM.} \label{table:num}
    \vspace{2mm}
 \begin{tabular*}{14cm}{ @{\extracolsep{\fill}} l c c c c}
    \hline
  &Energy (MeV) &Point matter (fm) &Point proton (fm) &Point neutron (fm)\\
    \hline
$^6$He ($0^+$)  &-28.37    &$2.71$                 &$2.03$                &$2.99$\\
$^6$Li   ($1^+$)  &-30.92    &$2.65$                 &$2.66$                &$2.65$\\
$^6$Li   ($3^+$)  &-29.87    &$2.42$                 &$2.43$                &$2.41$\\
$^6$Li   ($0^+$)  &-27.58    &$2.79$                 &$2.81$                &$2.77$\\
    \hline
  \end{tabular*}
  \end{center}
\end{table*} 
The experimental data and the corresponding results in the referenced
works~\cite{Itagaki2003,Furumoto2018} are also included for comparison. It should be noted that we
are using the same Hamiltonian and the same form of the basis wave functions. In
Fig.~\ref{fig:enespec}, it clearly shows that our REM method provides the almost consistent 
results for the $0^+$ states of $^6$He and $^6$Li nuclei as the references. Besides, for the $1^+$
ground state and the $3^+$ excited state of $^6$Li, the wave functions from our REM procedure
provide better results than the reference work, which means that we have found more sufficient wave
function through the evolution with the EOM. These results support the validity of the
REM. Furthermore, it should be noted that we are using one ensemble of the basis for both of the
$^6$He and $^6$Li calculations, it is interesting that one EOM can reproduce both the $\text{T}=0$
states and $\text{T}=1$ states, and it indicates that the REM may have the potential for the
investigation of the isospin mixing states in the future study. 

Next we shall check the accuracy of our calculations. We show the energy convergences with the
increasing number of basis in Fig.~\ref{fig:enecon}. It shows that the huge number of the basis have
been included and the binding energies of all these states are well converged. These results prove
that the number of basis in our calculations is sufficient to converge the energy
results. Furthermore, we can see that the converged results of $1^+$ and $3^+$ states in our
calculation are much lower than the results from the reference works. It denotes that the REM
procedure have found more effective basis, which should be included to the total wave function. 

It is also an essential topic to investigate the halo property of the $^6$He nucleus as well as the
$^6$Li nucleus. The $0^+$ ground state of $^6$He is the well known two-neutron halo. Likewise, the
$0^+$ excited states of $^6$Li also has the controversial halo 
property~\cite{Li2002}. To investigate the halo property in these two nuclei, we calculate the
root-mean-square (r.m.s.) radii of $^6$He and $^6$Li with the wave function from REM. The
corresponding results are shown in Fig.~\ref{fig:rmscon}.

In the left panel of this figure, the calculated r.m.s. radii of point matter, point proton and
point neutron of the $0^+$ state of $^6$He are $2.71$ fm, $2.03$ fm, and $2.99$ fm,
respectively. These results are showing the explicit halo property of the ground state of
$^6$He. From the right panel of Fig.~\ref{fig:rmscon}), one can also find that the r.m.s. radius of
point matter of the $0^+$ state of $^6$Li ($2.79$ fm) is larger than the radii of its $1^+$ ($2.65$
fm) and $3^+$ ($2.42$ fm). It implies that the $0^+$ state of $^6$Li can be treated as a halo state,
which is consistent with the experimental conclusion~\cite{Li2002}. These results show that the halo
property of these states can be naturally included in the ensemble of the basis from the
REM. Comparing with the reference works, we notice that our results on the r.m.s. radii are larger
than the results in the reference works, which are denoted by the dotted lines in
Fig.~\ref{fig:rmscon}. It indicates that our ensemble of basis from REM includes the basis, where
valence nucleons spread far from the core, so that we provide more dilute structure for the halo
states of $^6$He and $^6$Li nuclei than theirs. 

In the end, the detailed numerical results are summarized in Table~\ref{table:num}. The current
results should be the most accuracy calculation on $^6$He and $^6$Li nuclei within the GCM
framework. 

\section{Conclusion}
\label{sec:conclusion}

We perform the calculations for $^6$He and $^6$Li nuclei with a recently developed model named REM,
which can generate the ergodic ensemble of the basis wave functions. During this work, we generate
the basis wave functions from the procedure of REM and superpose them to construct the total wave
functions. The converged results for the energy and the r.m.s. radius of the $0^+$ state of $^6$He
as well as the $1^+$, $0^+$ and $3^+$ states of $^6$Li nuclei have been obtained in this work. The
halo properties of $^6$He and $^6$Li are well described in the current work, which indicates that
the REM can search the basis more efficiently. The current works on $^6$He and $^6$Li nuclei could
be the most accurate calculations within the GCM framework to date. The benchmark calculations
performed in this work can be instructive for further calculation with REM.  

\begin{acknowledgement}
One of the author (M.K.) acknowledges that this work was supported by the JSPS KAKENHI Grant No. 19K03859
and by the COREnet program at RCNP Osaka University.
\end{acknowledgement}
\bibliographystyle{spphys}       
\bibliography{6He}

\end{document}